\documentclass[a4paper]{article}

\usepackage{INTERSPEECH2020}
\usepackage[activate]{microtype}
\usepackage{color}
\usepackage{balance}
\usepackage{pifont}
\usepackage{subfig}
\usepackage{bbm}
\usepackage{cite}

\title{Semi-Supervised Learning with Data Augmentation for End-to-End ASR}
\name{Felix Weninger$^1$, Franco Mana$^2$, Roberto Gemello$^2$, Jes\'{u}s Andr\'{e}s-Ferrer$^3$, Puming Zhan$^1$}
\address{
  Nuance Communications, Inc., $^1$Burlington, MA, USA, $^2$Torino, Italy, $^3$Valencia, Spain
  }
\email{\{felix.weninger,franco.mana,roberto.gemello,jesusandres.ferrer,puming.zhan\}@nuance.com}

\begin{document}

\maketitle
\begin{abstract}
In this paper, we apply Semi-Supervised Learning (SSL) along with Data Augmentation (DA) for improving the accuracy of End-to-End ASR. 
We focus on the consistency regularization principle, which has been successfully applied to image classification tasks, and present sequence-to-sequence (seq2seq) versions of the FixMatch and Noisy Student algorithms. 
Specifically, we generate the pseudo labels for the unlabeled data on-the-fly with a seq2seq model after perturbing the input features with DA. 
We also propose soft label variants of both algorithms to cope with pseudo label errors, showing further performance improvements.
We conduct SSL experiments on a conversational speech data set (doctor-patient conversations) with 1.9\,kh manually transcribed training data, using only 25\,\% of the original labels (475\,h labeled data).
In the result, the Noisy Student algorithm with soft labels and consistency regularization achieves 10.4\,\% word error rate (WER) reduction when adding 475\,h of unlabeled data, 
corresponding to a recovery rate of 92\,\%.
Furthermore, when iteratively adding 950\,h more unlabeled data, our best SSL performance is within 5\,\% WER increase compared to using the full labeled training set (recovery rate: 78\,\%).


\end{abstract}
\noindent\textbf{Index Terms}: automatic speech recognition, semi-supervised learning, data augmentation, sequence-to-sequence, end-to-end

\section{Introduction}
End-to-end (E2E) systems have become a focus of ASR research in recent years, due to their ability of integrating all components of an ASR system in a single deep neural network (DNN), which greatly simplifies and unifies the training and decoding process \cite{Graves2012-STW,Bahdanau2015-ETE,Battenberg2017-ENT,Prabhavalkar2017-ACO,Li2018-AAT}.
The Sequence-to-Sequence (seq2seq) model with attention is one of the model architectures for 
E2E ASR systems which has shown state-of-the-art performance 
\cite{Chan2016-LAA,Chiu2018-SOT,Zeyer2018-ITO,Weninger2019-LAS,Tuske2020-SHA}.
However, a general observation is that E2E ASR needs large amounts of training data for achieving state-of-the-art performance, especially when no language model trained with external text data is included in the system. 
Data Augmentation (DA) and semi-supervised learning (SSL)
are two approaches that can be used for improving E2E model performance with limited amounts of manually transcribed training data.

DA perturbs (usually randomly) the input data without altering the corresponding labels. This not only increases the variety of the data, but also serves as implicit regularization to avoid overfitting \cite{Hernandez2018-DAI}.
It has been successfully used for both conventional \cite{Ko2015-AAF,Cui2015-DAF,Zhou2020-TRA} and E2E ASR systems \cite{Saon2019-SNI,Nguyen2020-IST}.
In particular, the SpecAugment approach proposed in \cite{Park2019-SAA} has shown impressive improvement for seq2seq based E2E ASR models.

SSL (also called semi-supervised training) aims at leveraging unlabeled data for improving ASR model accuracy.
In the self-training paradigm for SSL, a seed model trained with limited amount of labeled data is used to generate transcriptions (pseudo labels) for unlabeled data (cf.\ \cite{Lamel2002-LSA}). This procedure can be iterated on additional unlabeled data \cite{Khonglah2020-ISS}.
Another possible implementation of SSL is via teacher-student training, where a `student' model is trained to replicate the outputs of a powerful `teacher' model on the unlabeled data \cite{Gibson2017-SST}.

The central research question of our paper is how to best integrate SSL with DA for training E2E ASR systems.
So far, the use of DA with SSL for E2E ASR has been largely limited 
to a simple cascade of both, i.e., doing pseudo label generation for the unlabeled data and then applying DA
\cite{Synnaeve2019-ETE,Kahn2019-STF,Huang2019-SST,Chen2020-SSA}.
In contrast, for image classification, several algorithms have recently been proposed that use DA for teacher-student training \cite{Xie2019-STW} and \emph{consistency regularization} in SSL \cite{Xie2019-UDA,Berthelot2019-MAH,Sohn2020-FSS}.
Consistency regularization stems from the intuition that a small perturbation to an input data sample should not change the output distribution a lot.
However, these SSL algorithms were designed only for static classification and need to be modified to support the seq2seq ASR use case. 

In this regard, our paper makes the following contributions:
First, we modify the Noisy Student \cite{Xie2019-STW} and FixMatch \cite{Sohn2020-FSS} algorithms -- which have only been applied to image classification so far -- for the seq2seq ASR use case.
Second, we show performance improvements for both algorithms by using soft labels and consistency training (via SpecAugment and dropout).
Finally, we demonstrate additional gains from iterative  generation of pseudo labels by exploiting a larger amount of unlabeled data. We show that our proposed methods outperform the simple approach of doing DA after generating pseudo labels.

\textbf{Relation to prior work:}
Several studies have recently investigated SSL techniques for E2E ASR, e.g.\ representation learning \cite{Ling2019-DCA}, the usage of external text data \cite{Karita2018-SSE}, text-to-speech \cite{Karita2019-SSE}, and transcriptions generated by conventional ASR \cite{Li2019-SST}. 
Regarding self-training for E2E ASR, \cite{Kahn2019-STF} proposed data filtering and ensemble schemes, generating hard pseudo labels via beam search.
In \cite{Dey2019-ESS}, dropout was employed to improve the pseudo label accuracy and confidence measure, due to its well-known model ensembling property.
All of these works did not consider the interaction of SSL with DA, which we investigate in our study.
Very recently, \cite{Masumura2020-SLC} proposed teacher-student learning with DA for consistency training, without considering the Noisy Student or the FixMatch algorithm or soft labels as in our work.

\section{Methods}

\label{sec:methodology}

For our E2E ASR models, we use an encoder-decoder architecture with attention as described in \cite{Weninger2019-LAS}, which is similar to Listen-Attend-Spell (LAS) \cite{Chan2016-LAA}.
The ASR task is treated as a seq2seq learning problem:
The model $M$ is trained to predict a sequence $y_j$ of symbols (here, we use sub-word units) from a sequence of acoustic features (usually a spectrogram $x$).

The encoder $e$ creates a hidden representation 
of the acoustic features.
Here, $e$ is implemented as a stack of convolutional (CNN) layers followed by bidirectional Long Short-Term Memory (bLSTM) layers.
The decoder is similar to an RNN LM that takes into account a context vector $c_j$. 
Bahdanau attention \cite{Bahdanau2015-NMT} is used to focus $c_j$ on various parts of the encoder output. The output distribution $p_j = p(y_j | y_1, \dots, y_{j-1}, x) = M(x,y_{<j})$ for the $j$-th symbol is computed by
\begin{eqnarray}
    s_j &=& f(s_{j-1}, \text{Embedding}(y_{j-1}), c_{j-1}) , \label{eq:decoder_state} \\
    c_j &=& \text{Attention}(s_{j}, e(x)) , \\    
    s_j' &=& \text{Dense}(s_j, c_j), \\
    p_j &=& g(s_j'), 
    \label{eq:softmax}
\end{eqnarray}
where $g$ is a softmax layer and $f$ is a stack of LSTM layers.

To perform ASR using the seq2seq model, hypotheses are generated by beam search.
The model is conditioned on its previous output $y_{j-1}$ in Eq.~(\ref{eq:decoder_state}).
The sequence $y_1, \dots, y_j, \dots$ of output symbols is produced one at a time, until the end-of-sentence symbol is reached. 

\subsection{Supervised and Semi-Supervised Training}
For \emph{supervised} training, the following cross-entropy (CE) loss is minimized for a mini-batch $\mathcal{B}$ of training examples:
\begin{equation}
    \mathcal{L}_\theta(\mathcal{B}) = \frac{1}{|\mathcal{B}|} \sum_{i \in \mathcal{B},j} \mathcal{L}^{\text{CE}}(y^*_{i,j},M_\theta(A(x_i), y^*_{i,<j})) .
    \label{eq:loss}
\end{equation}
Here, $y^*_{i,j}$ is the $j$-th symbol in the ground truth (GT) transcription of the $i$-th utterance in $\mathcal{B}$, represented as a one-hot vector,
$A(x_i)$ denotes the features of the $i$-th utterance in $\mathcal{B}$ after data augmentation (cf.\ Section \ref{sec:DA}), and $\theta$ are the model parameters.

In \emph{semi-supervised} training, the loss \eqref{eq:loss} is split into a labeled and an unlabeled part as $\mathcal{L}_\theta(\mathcal{B}) = (1/|\mathcal{B}|) ( \mathcal{L}_\theta(\mathcal{B}^l) + \mathcal{L}_\theta(\mathcal{B}^u))$,
\begin{align}
    \mathcal{L}_\theta(\mathcal{B}^l) =& 
    \sum_{i \in \mathcal{B}^l,j} \mathcal{L}^{\text{CE}}(y^*_{i,j},M_\theta(A(x^l_i), y^*_{i,<j})), \\
    \mathcal{L}_\theta(\mathcal{B}^u) =& 
    \sum_{i \in \mathcal{B}^u,j} \mathbbm{1}({\hat{p}_{i,j} \geq C}) \, \mathcal{L}^{\text{CE}}(\hat{y}_{i,j},M_\theta(A(x^u_i), \hat{y}_{i,<j})) ,
    \label{eq:unlab_loss}
\end{align}
where $\mathcal{B}^l$ contains labeled and $\mathcal{B}^u$ contains unlabeled utterances, $\mathcal{B} = \mathcal{B}^l\,\cup\,\mathcal{B}^u$.
For the unlabeled loss \eqref{eq:unlab_loss}, pseudo labels, i.e.\ pseudo truth (PT) transcriptions $\hat{y}_{i,j}$, need to be obtained (cf.\ Section \ref{sec:PT}). 
Denoting the pseudo label confidence for utterance $i$ and token $j$ as $\hat{p}_{i,j}$,
confidence-based filtering is included in the loss \eqref{eq:unlab_loss} by setting a confidence threshold $C > 0$.

\subsection{Data Augmentation (DA)}

\label{sec:DA}

In our training framework, DA is modeled by a function $A$, which also depends on the current state of the random number generator.
We parameterize the function $A$ based on SpecAugment (SA) \cite{Park2019-SAA} with hyperparameters $F_\text{max}$, $T_\text{max}$, $m_F$ and $m_T$. 
The SA algorithm masks (replaces by zeros) up to $F_\text{max}$ contiguous frequency bands and up to $T_\text{max}$ contiguous time frames in the  spectrogram $x$.
The starting positions and the actual number of masked rows / columns are sampled from a uniform distribution.
This process is repeated $m_F$ times for masking frequencies and $m_T$ times for masking time frames.
Due to the combinatorial explosion of possible corruptions applied by $A$ to a single input $x$, the usage of SA results in a practically infinite amount of training data.

\subsection{Pseudo Labeling}

\label{sec:PT}
 
We obtain pseudo labels for SSL based on PT transcriptions of the unlabeled data.
These are generated `offline' prior to training using beam search with a seq2seq model.
Data augmentation is turned off during this offline PT generation phase.
Since the PT generation is done only once, it is reasonable to use a large beam size $W$ to improve the PT quality (here, we set $W=16$).
Furthermore, we apply a heuristic loop filtering technique to the PT utterances similar to the one from \cite{Kahn2019-STF}.
This helps us avoid reinforcing the well-known `looping' problem, where the seq2seq model keeps repeating the same n-gram.

Semi-supervised training can be performed by directly using the PT transcriptions as pseudo labels $\hat{y}_{i,j}$ in \eqref{eq:unlab_loss}. 
Alternatively, we can dynamically update the PT transcriptions in the training process as in the FixMatch or Noisy Student approaches.

\subsubsection{DA for Consistency Training (FixMatch)}

\begin{figure}
    \centering
    \includegraphics[width=0.8\columnwidth]{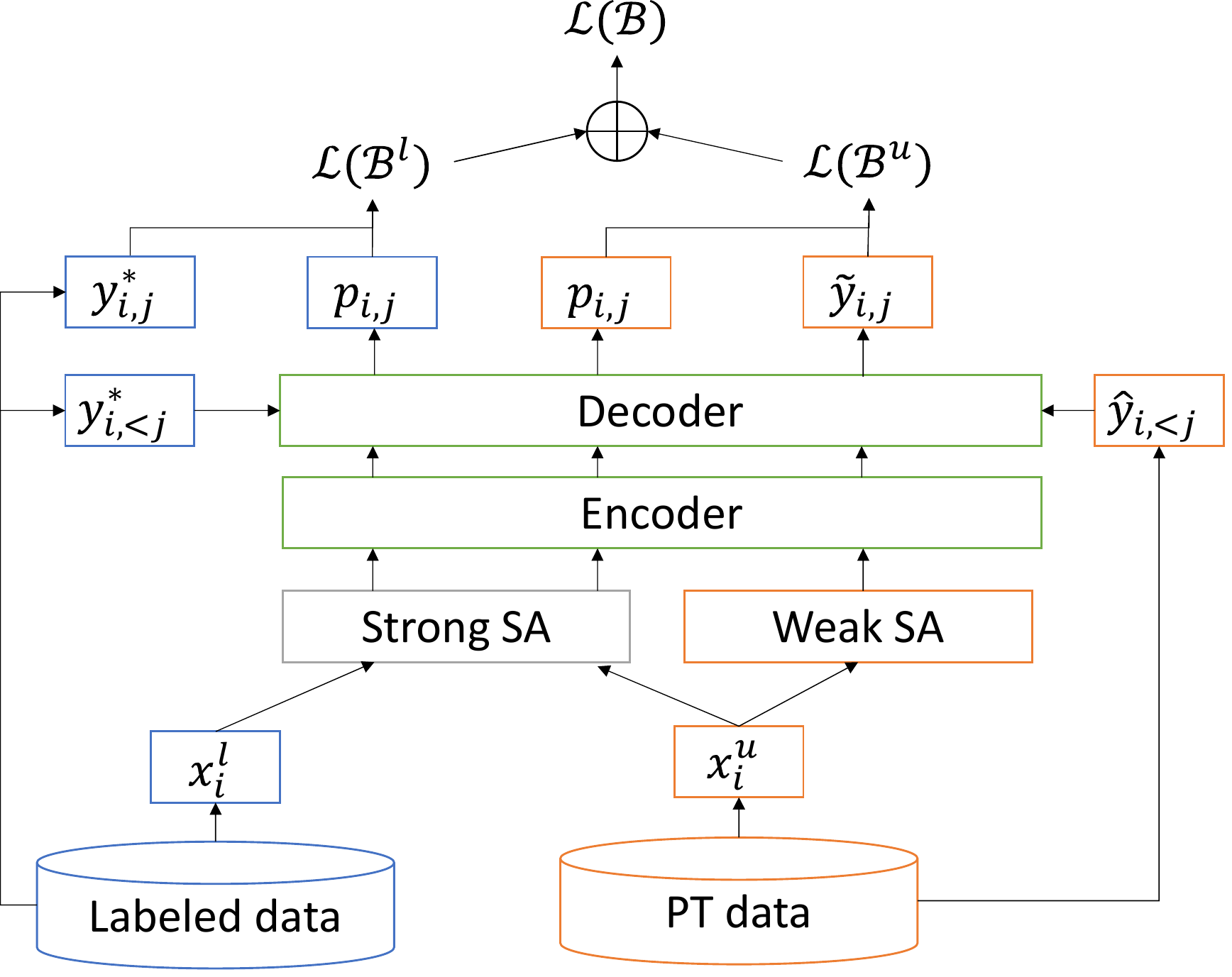}
    \caption{Sequence-to-sequence FixMatch algorithm.}
    \label{fig:fm_alg}
\end{figure}

The FixMatch algorithm is a self-training method that was proposed in \cite{Sohn2020-FSS} for image classification.
The method implements \emph{consistency training} by applying two kinds of data augmentation to the input $x$ in both  the unlabeled loss and the pseudo label generation, thus encouraging the outputs to be consistent for both augmented inputs.
Here, we present an extension of FixMatch to seq2seq models.
Specifically, we generate pseudo labels $\tilde{y}_{i,j}$ for unlabeled training examples $x^u_i$ as:
\begin{equation}
\tilde{y}_{i,j} = M_\theta(A_w(x^u_i), \hat{y}_{i,<j}) ,
\label{eq:fixmatch_pseudo}
\end{equation}
where $A_w$ is a \emph{weak} augmentation function.
We then use $\tilde{y}_{i,j}$ in place of the PT label $\hat{y}_{i,j}$ in Eq.~(\ref{eq:unlab_loss}).
\figurename~\ref{fig:fm_alg} depicts our FixMatch algorithm for seq2seq ASR.

As argued by \cite{Sohn2020-FSS}, choosing a weak augmentation $A_w$ for pseudo label generation instead of the `strong' augmentation $A$ in the unlabeled loss ensures reasonable accuracy of the pseudo labels and improves the convergence.
Moreover, our experiments show that applying the confidence threshold $C$ leads to gradual annealing of the supervised training signal, as the model becomes more and more confident and hence includes more and more unlabeled data as the training proceeds.
Finally, by applying dropout in all forward passes of the model (including the pseudo label generation), we add further regularization to the training.

Besides extending the algorithm to seq2seq, we make two further modifications. 
First, we also investigate using soft labels, in order to make the training more robust against pseudo label errors in case of ambiguous instances.
Second, while \cite{Sohn2020-FSS} applied weak augmentation on labeled data, we found strong augmentation to perform better in practice.

\subsubsection{Teacher-Student Training with DA (Noisy Student)}

\begin{figure}
    \centering
    \includegraphics[width=0.8\columnwidth]{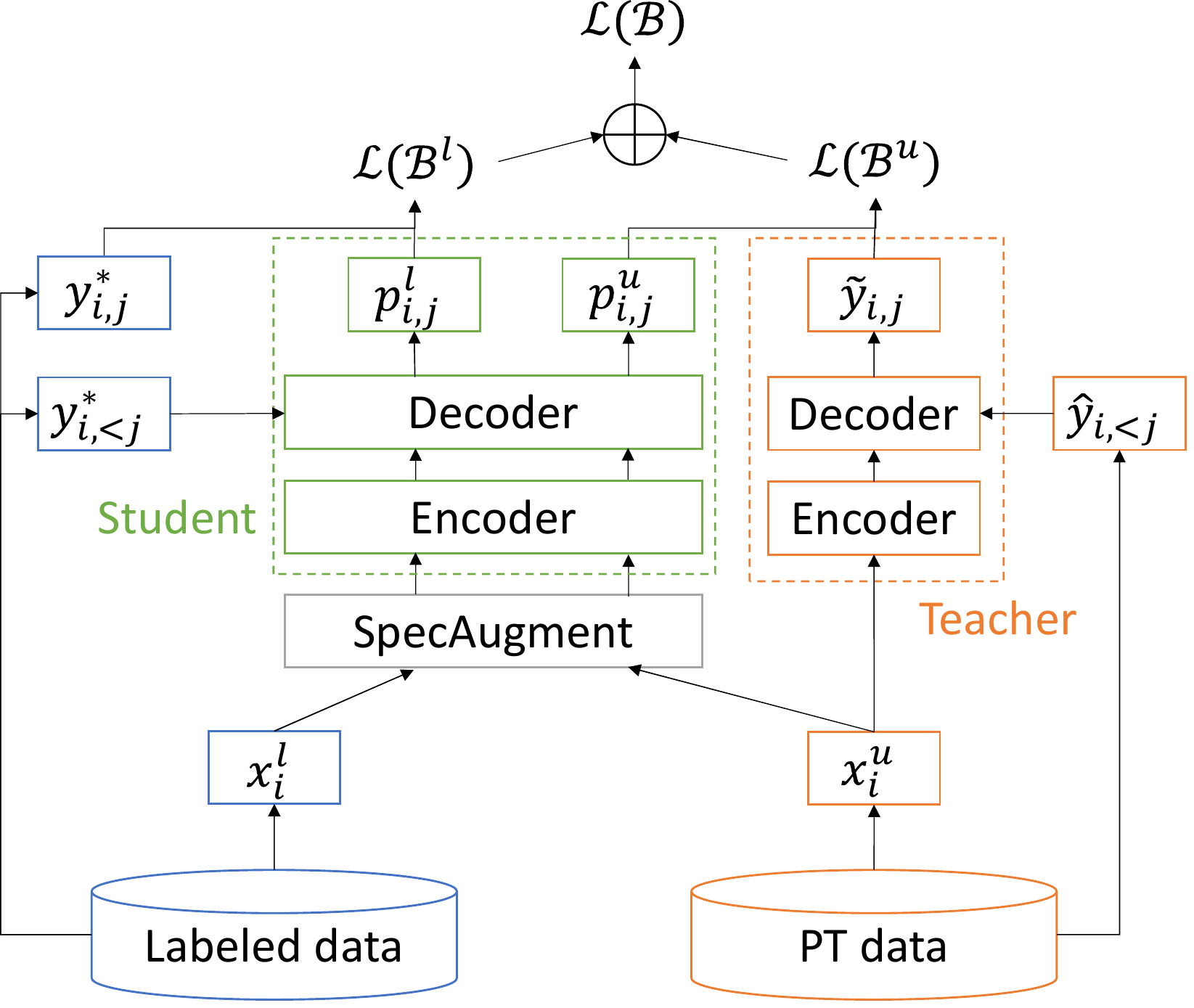}
    \caption{Sequence-to-sequence Noisy Student algorithm.}
    \label{fig:noisy_student_alg}
    \vspace{-5mm}
\end{figure}

The Noisy Student algorithm \cite{Xie2019-STW} is another recent SSL algorithm proposed for image classification, combining teacher-student training with DA. 
The main difference to FixMatch is that pseudo labels are generated by a pretrained teacher model $M_{\theta^*}$ with `frozen' parameters $\theta^*$, instead of the current model parameters $\theta$: 
\begin{equation}
\tilde{y}_{i,j} = M_{\theta^*}(x^u_i, \hat{y}_{i,<j}) .
\end{equation}
In our experiments, the teacher model is of the same topology as the student model. 
\figurename~\ref{fig:noisy_student_alg} depicts our Noisy Student algorithm for ASR.

Since both hard and soft teacher labels (distillation) were considered in \cite{Xie2019-STW}, we also explored both for the E2E ASR task.
The hard label variant is equivalent to using the PT transcriptions in the unlabeled loss \eqref{eq:unlab_loss}, i.e.\ $\tilde{y}_{i,j} = \hat{y}_{i,j}$.
For the soft label variant, we (re-)compute soft labels on-the-fly via a forward pass on the teacher model.
This is done to avoid storing full posteriors, which is hard for large label spaces and PT data sets.
The on-the-fly label computation also allows us to include dynamic modifications of the input and the model. 
Specifically, we explore consistency training for the Noisy Student algorithm by (weakly) augmenting the input for pseudo label generation as in FixMatch (Eq.~\ref{eq:fixmatch_pseudo}), or running the teacher model with dropout included.
Our Noisy Student framework also subsumes generating hard labels with consistency training, similar to \cite{Masumura2020-SLC}.

\subsubsection{Iterative SSL}

We also investigate iterative SSL, i.e., several rounds of PT generation by the E2E model via beam search, in two ways.
First, we perform several iterations of the Noisy Student algorithm similar to \cite{Xie2019-STW}.
Second, we also compare to an iterative self-training algorithm with DA inspired by \cite{Chen2020-SSA}, that regenerates PT transcription for each mini-batch based on the current model parameters. 
However, while \cite{Chen2020-SSA} used greedy search, we found this to yield insufficient PT quality and thus adopted a lightweight beam search ($W=4$).

\subsection{Decoding}

The decoding loss is extended with several terms so as to avoid the bias of  seq2seq models to both deletions and insertions. 
Specifically, we maximize the following score via histogram pruning target synchronous beam search:
\begin{equation}
  \log p(y\!\mid\! x)   + \lambda_{\text{cov}} \operatorname{cov}(x,y)  
  + \lambda_{\text{wip}} |y|
  + \left(\frac{5 + |u|}{5+1}\right)^{\lambda_{\text{rlp}}}
\label{eq:infer}
\end{equation}
where $p(y | x)$ is the model probability; $\operatorname{cov}(x,y)$ is a coverage score (see Eq.~(11) of~\cite{Chorowski2017-TBD}) weighted by $\lambda_{\text{cov}}$ which adds a bonus per each encoder states that is attended more than 0.5 while decoding;   $\lambda_{\text{wip}}$ is a constant word insertion penalty  to balance the average probability reduction per decoded token; and the last term  is a root length bonus/penalty to better approximate the length bias~\cite{Wu2016-GNM}.

\begin{figure}[t]
    \centering
    \includegraphics[width=.85\linewidth]{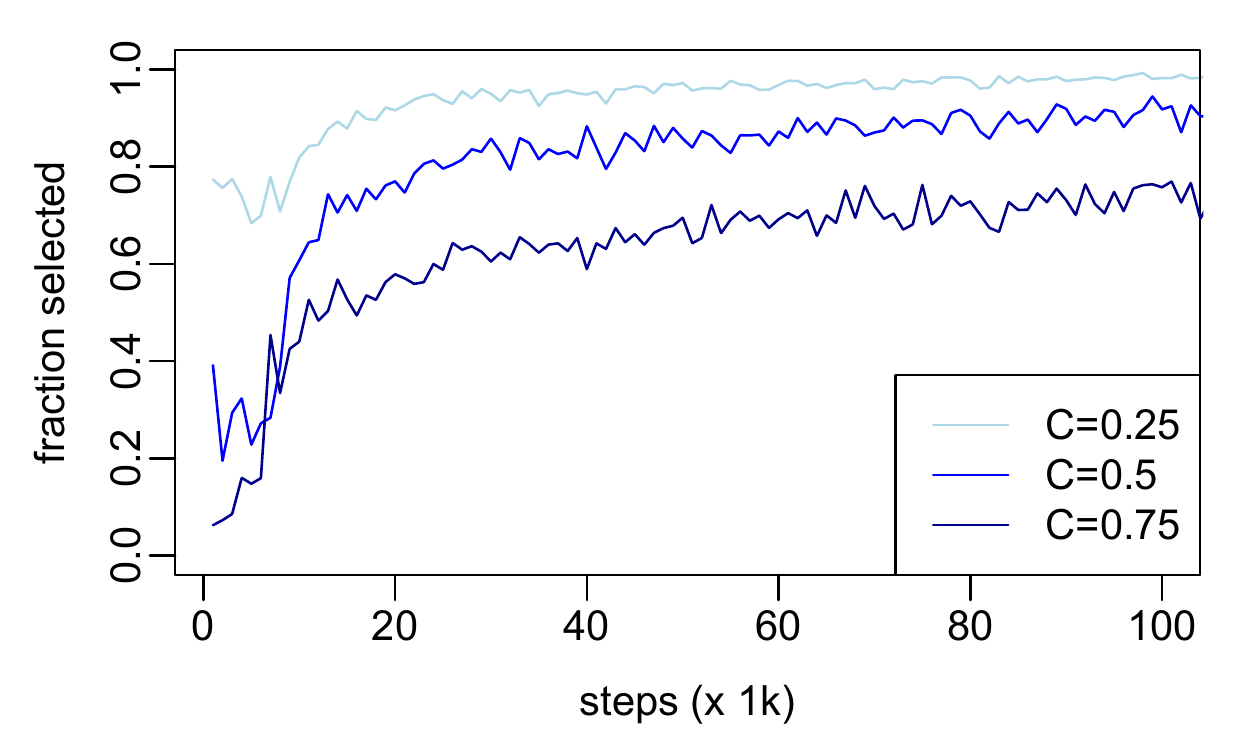}
    \\[-1mm]
    \includegraphics[width=.85\linewidth]{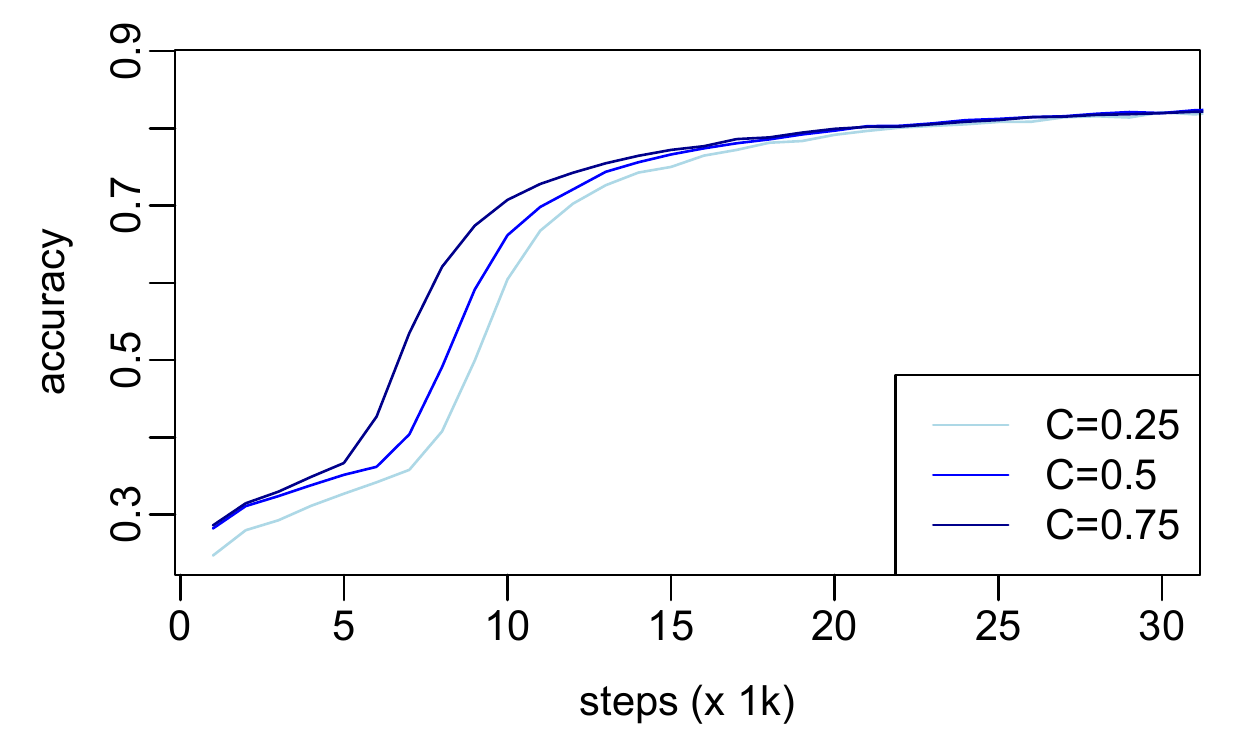}
\caption{FixMatch behavior over time for various confidence thresholds $C$. 
}
    \label{fig:fm_acc}
    \vspace{-5mm}
\end{figure}

\section{Experiments}

\label{sec:experiments}

\subsection{Data Set}

Our methods are evaluated on a conversational speech transcription task (doctor-patient conversations).
All speech data are anonymized field data.
Our experiments are based on a corpus of 1.9\,k hours manually end-pointed and transcribed speech.
We divided the corpus into four parts of equal size, treating only the first one (475\,h) as labeled.
In a first set of experiments, we treated the second part (475\,h) as unlabeled data (ignoring the labels).
We performed additional experiments adding the remaining 950\,h of data as unlabeled data.
In all cases, we measure the word error rate (WER) on a test set of 300\,k words (26.8\,h).

\subsection{E2E Model Configuration}

Our E2E models use 80-dimensional log Mel filterbank outputs as input features.
The inputs are first passed through a stack of 3 CNN layers, which is parameterized so as to yield a 512-dimensional embedding for each input frame.
Subsequently, there is a pyramid bLSTM encoder with 6 layers (512 LSTM units per layer and direction), which performs frame decimation (by a factor of 2) after every other layer, thus reducing the frame rate by a factor of 8.
The decoder uses 2 (unidirectional) LSTM layers (1\,024 units per layer).
The softmax output layer predicts the posterior probabilities for 2\,k word piece targets determined on the training set. 
In total, the model has 66\,m parameters.

\subsection{Training Parameters}

The E2E models are trained with 
dropout \cite{Srivastava2014-DAS} (with probability 0.3), label smoothing for hard labels \cite{Szegedy2016-RTI} (with probability 0.9 for the target class), and early stopping (using a validation set held out from the training data) in order to improve generalization.
The SpecAugment parameters are set as $F_\text{max}=35$, $T_\text{max}=50$, $m_F=1$, $m_T=2$.
For the weak augmentation $A_w$ in pseudo label generation, SA is parameterized with $F_\text{max}=5$, $m_F=1$, $T_\text{max}=m_T=0$ (frequency masking only).

\section{Results}

\label{sec:results}

\subsection{Supervised Training}

First, we performed baseline experiments with supervised training and SA. 
Using 475\,h GT data, we obtain 16.77\,\% WER (cf.\ \tablename~\ref{tab:fm_results}), which is the baseline for our SSL experiments. 
Using 1.9\,kh GT data, we achieve 13.79\,\% WER, which is the lower bound on the WER attainable with SSL.

\begin{table}[t!!]
    \caption{FixMatch results using 475\,h labeled data (GT), 475\,h unlabeled data (PT) and 200\,k training steps.}
    \label{tab:fm_results}
    \centering
    \begin{tabular}{lllll}
    $C$ & PT labels & PT noise & Init & WER \\ 
     \hline\hline
    \multicolumn{5}{c}{\em Supervised (475h)} \\
    -- & -- & -- & random & 16.77 \\
    \hline\hline
    \multicolumn{5}{c}{\em Semi-supervised, FixMatch} \\
    0.5 & hard & dropout & random & 15.97 \\
    0.5 & hard & + SA ($F=5$) & random & 16.13 \\
    0.5 & hard & + strong SA & random & 16.34 \\
    \hline
    0.5 & soft & dropout & random & 15.79 \\
    \hline
    0.0 & soft & dropout & 475\,h GT & \bf 15.22 \\
    \hline
    \end{tabular}
    \vspace{-3mm}
\end{table}

\subsection{FixMatch}

\figurename~\ref{fig:fm_acc} shows the behavior of the FixMatch algorithm for different confidence thresholds $C$. 
It can be seen that the fraction of selected tokens decreases significantly with increasing $C$, thus effectively training on less data.
However, as training progresses, the validation accuracy varies little between different $C$, which also results in similar WER.
\tablename~\ref{tab:fm_results} shows the ASR performance for various FixMatch settings, using a short training schedule of 200\,k maximum steps to reduce experimental turnaround time.
We observe that adding SA-based consistency training on top of dropout does not give an improvement; however, it is confirmed that using strong augmentation instead of weak augmentation in pseudo label computation degrades the results.
The usage of soft labels leads to a slight gain (1.1\,\% WERR, where WERR is defined as relative WER reduction).

To further improve on the FixMatch performance, we initialized the FixMatch parameters from the model previously trained on 475\,h GT data, while also dispensing with the confidence-based selection, since in this case we assume that the PT labels are of high quality from the start.
This approach led to a significant improvement (3.6\,\% WERR), yielding 9.2\,\% WERR from FixMatch compared to the supervised baseline.

\begin{table}[t]
    \caption{Comparison of FixMatch and Noisy Student methods, using 475\,h of labeled (GT) 
    and 475\,h of unlabeled data (PT), and 70 training epochs.
    SA: SpecAugment.
    Init(ialization): random or from the supervised training on 475\,h GT.
    }
    \label{tab:2gpu_results}
    \centering
    \begin{tabular}{lllll}
    PT labels & PT noise & Init & WER & WRR \\ 
    \hline\hline
    \multicolumn{5}{c}{\em Semi-supervised, Noisy Student} \\
    hard & none & random & 15.60 & 61.6 \\
    hard & SA ($F=5$) & random & 15.26 & 79.5 \\
    \hline
    soft & none & random & 15.16 & 84.7 \\
    soft & dropout & random & 15.10 & 87.9 \\
    soft & SA ($F=5$) & random & \bf 15.02 & \bf 92.1 \\
    \hline\hline
    \multicolumn{5}{c}{\em Semi-supervised, FixMatch} \\
    soft & dropout & 475\,h GT & 15.04 & 91.1 \\
    \hline\hline
    \multicolumn{5}{c}{\emph{Semi-supervised, iterative self-training} \cite{Chen2020-SSA}} \\
    hard & none & 475\,h GT & 15.34 & 75.3 \\
    \hline\hline
    \multicolumn{5}{c}{\em Supervised (950\,h), oracle performance} \\
    -- & -- & random & \em 14.87 & \em 100.0 \\
    \hline
    \end{tabular}
\end{table}

\begin{table}[t!!]
    \caption{
    Results with iterative PT generation using 
    a total of 475\,h labeled and 1.4\,kh unlabeled data.
    }
    \label{tab:results_1.4kh}
    \centering
    \begin{tabular}{lllll}
    PT labels & PT noise & Init & WER & WRR \\ 
    \hline\hline
    \multicolumn{5}{c}{\em Semi-supervised, Noisy Student} \\
    hard & none & random & 14.88 & 63.4 \\
    \hline
    soft & none & random & 14.52 & 75.5 \\
    soft & SA ($F=5$) & random & \bf 14.44 & \bf 78.2 \\
    \hline\hline
    \multicolumn{5}{c}{\em Supervised (1.9\,kh), oracle performance} \\
    -- & -- & random & \em 13.79 & \em 100.0 \\
    \hline
    \end{tabular}
\end{table}

\subsection{Noisy Student}

Results of the Noisy Student algorithm are shown in \tablename~\ref{tab:2gpu_results}, using a training schedule of 70 epochs (with early stopping).
We observe 2.2\,\% WERR from using consistency regularization compared to the hard label Noisy Student algorithm.
Moreover, the usage of soft labels improves the results considerably (2.8\,\% WERR).
The best result (15.02\,\% WER) is obtained by combining soft labels and consistency training.
The best Noisy Student setup performs similar to the best FixMatch setup, and outperforms the iterative self-training similar to \cite{Chen2020-SSA} by 2\,\%. 

We also assess the WER recovery rate (WRR) of our SSL algorithms as defined in \cite{Kahn2019-STF}, which is the ratio of performance gain achieved by adding unlabeled data vs.\ the gain from adding the same amount of \emph{labeled} data (in this case, training with 950\,h labeled data).
The simplest variant, which uses one-shot hard label PT generation (15.60\,\% WER), achieves 61.6\,\% WRR.
The other SSL methods in Table 2, which all use on-the-fly PT label generation, can improve on this baseline.
In particular, our variant using soft labels along with consistency training obtains 92.1\,\% WRR. 

While we found only minor performance gains from multiple iterations of the Noisy Student algorithm on the {\em same} unlabeled data, we obtained additional improvements when including additional 950\,h PT data, for a total of 1.4\,kh PT data.
In this case, the PT labels $\hat{y}_{i,j}$ were generated using a model trained on 475\,h GT and 475\,h PT using the Noisy Student algorithm.
Results are shown in \tablename~\ref{tab:results_1.4kh}.
With soft PT labels and consistency training, 
we achieve a 4\,\% improvement over the previous best result.

\section{Conclusions}

\label{sec:conclusions}

In this paper, we presented a comprehensive study on SSL strategies for end-to-end ASR. 
We investigated the FixMatch and Noisy Student algorithms for ASR and demonstrated improvements from using soft labels and consistency training.
We believe that this is due to their ability to reduce error reinforcement in SSL. 
In the result, the performance of SSL can approach the one of supervised learning with similar amounts of data.

In future work, we will extend our investigation to different DA schemes beyond SpecAugment (e.g.\ \cite{Saon2019-SNI},\cite{Nguyen2020-IST}). We will also look at  iteratively including more and more unlabeled data while increasing the model size as in \cite{Xie2019-STW}.

\bibliographystyle{IEEEtran}

\balance

\bibliography{mybib}

\end{document}